\documentclass[aps,prb,showpacs,twocolumn,amssymb,floats,epsfig]{revtex4}
\usepackage{graphicx}
\usepackage{amsmath}
\usepackage{mathbbol}
\usepackage{relsize}
\usepackage{afterpage}
\usepackage{braket}
\usepackage{hyperref}
\usepackage{subfigure}
\usepackage[usenames,dvipsnames]{color}

\newcommand{\ct}{\cite}
\newcommand{\bi}{\bibitem}
\newcommand{\be}{\begin{equation}}
\newcommand{\ee}{\end{equation}}
\newcommand{\beq}{\begin{equation}}
\newcommand{\eeq}{\end{equation}}
\newcommand{\bea}{\begin{eqnarray}}
\newcommand{\eea}{\end{eqnarray}}
\newcommand{\bitem}{\begin{itemize}}
\newcommand{\eitem}{\end{itemize}}
\newcommand{\al}{\alpha}
\newcommand{\de}{\delta}

\newcommand{\ga}{\gamma}

\newcommand{\si}{\sigma}

\newcommand{\dg}{\dagger}
\newcommand\vk{{\vec k}}
\newcommand\vn{{\vec n}}

\newcommand\ua{\uparrow}
\newcommand\da{\downarrow}
\newcommand{\non}{\nonumber}
\newcommand{\noi}{\noindent}

\begin{document}

\title{Effects of periodic kicking on dispersion and wave packet dynamics in 
graphene}
\author{Adhip Agarwala$^1$, Utso Bhattacharya$^2$, Amit Dutta$^2$, and 
Diptiman Sen$^3$}
\affiliation{\small{$^1$Department of Physics, Indian Institute of Science, 
Bengaluru 560012, India \\
$^2$Department of Physics, Indian Institute of Technology, Kanpur 208016, 
India \\
$^3$Centre for High Energy Physics, Indian Institute of Science, Bengaluru
560012, India}}

\begin{abstract}
We study the effects of $\de$-function periodic kicks on the Floquet 
energy-momentum dispersion in graphene. We find that a rich variety of 
dispersions can appear depending on the parameters of the kicking: at 
certain points in the Brillouin zone, the dispersion can become linear but 
anisotropic, linear in one direction and quadratic in the perpendicular 
direction, gapped with a quadratic dispersion, or completely flat (called 
dynamical localization). We show all these results analytically and 
demonstrate them numerically through the dynamics of wave packets propagating 
in graphene. We propose experimental methods for producing these effects.
\end{abstract}


\maketitle

\section{Introduction}
\label{sec_intro}

Quantum systems driven periodically in time have been extensively studied
in recent years from many points of view, such as the generation of defects
\ct{mukherjee08,mukherjee09}, coherent destruction of tunneling
\ct{grossmann91,kayanuma94} and dynamical freezing \ct{das10}, dynamical
saturation \ct{russomanno12} and localization \ct{alessio13,bukov14,nag14},
dynamical fidelity \ct{sharma14}, edge singularity in the probability 
distribution of work \ct{russomanno15} and thermalization \ct{lazarides14}
(see Ref.~\onlinecite{dutta15} for a review). These studies have become more 
important following the proposals of Floquet (irradiated) graphene 
\ct{gu11,kitagawa11,morell12}, Floquet topological insulators and the 
generation of topologically protected edge states \ct{kitagawa10,lindner11,
jiang11,trif12,gomez12,dora12,cayssol13,liu13,tong13,rudner13,katan13,
lindner13,kundu13,basti13,schmidt13,reynoso13,wu13,manisha13,perez1,perez2,
perez3,reichl14,manisha14}; some of these aspects have been experimentally 
studied \ct{kitagawa12,rechtsman13,puentes14}. We note that topologically 
ordered systems have also been studied using other driving schemes, for 
instance, sudden quenching; see Refs. 
\onlinecite{patel13,rajak14,sacramento14}.

Dynamical localization is a particularly interesting phenomenon; here the 
electrons become completely localized in space due to periodic driving of 
some parameter in the Hamiltonian. Examples of systems showing dynamical 
localization include driven two-level systems \ct{grossmann91}, classical and 
quantum kicked rotors \ct{chirikov81,ammann98}, the Kapitza pendulum 
\ct{kapitza51}, and bosons in an optical lattice \ct{horstmann07}.

There has been a tremendous amount of research on graphene in the last several 
years, both theoretical and experimental~\ct{been08,neto09,rev3,rev4,rev5}. 
Graphene is a two-dimensional hexagonal lattice of carbon atoms in which the 
$\pi$ electrons hop between nearest neighbors. The
spectrum is gapless at two points (called $\vec K$ and $\vec K'$) in
the Brillouin zone, and the energy-momentum dispersion around 
those points has the Dirac form $E_\vk = \hbar v_F |\vk|$, where $v_F
\simeq 10^6 m/s$ is the Fermi velocity. The Dirac nature of the electrons
gives rise to many interesting properties of this material. The ability
to manipulate the dispersion of electrons in graphene can therefore
be expected to give rise to some novel applications.

In this work, we will study the effects of periodic driving on an electron
moving on the graphene lattice. We will use Floquet theory 
to look at the stroboscopic properties (i.e., measured at the end of each time 
period of the driving) of the system. In particular we will be interested in
the form of the quasi-energy dispersion and the dynamics of wave packets
when various parameters in the Hamiltonian are given periodic $\de$-function
kicks. (We note here that the effects of periodic kicking on charge transport 
and optical conductivity of a graphene nanoribbon have been studied in 
Ref.~\onlinecite{babajanov14}. Further, a proposal for using periodic pulses 
to simulate the effects of curvature in graphene has been made in 
Ref.~\onlinecite{mishra15}). As we will see, the advantage of periodic 
$\de$-function kicks (in contrast to sinusoidal driving) is that the problem 
can be studied analytically to a large extent \ct{nag14,dasgupta14}. We will 
show that for certain specific values of the driving parameters, the 
quasi-energy is completely independent of the momentum; this leads to zero 
group velocity and therefore to dynamical localization of wave packets. We 
also find other interesting phenomena. For example, the dispersion can become 
gapless along a particular line in the momentum space which results in a 
movement of wave packets along one particular direction in real space. We can 
also make the dispersion linear in one direction and quadratic in the 
perpendicular direction (called a semi-Dirac dispersion), or gapped and 
quadratic in both directions around certain points in the Brillouin zone.

The plan of our paper is as follows. In Sec.~\ref{sec_graphene}, we recall 
the Hamiltonian and energy dispersion of graphene. In Sec.~\ref{sec_floquet}, 
we describe the ideas of periodic $\de$-function kicking of the Hamiltonian 
and the quasi-energy dispersion for the most general form of kicking. 
Sec.~\ref{sec_dynamics} describes how the 
dynamics of a Gaussian wave packet can be studied numerically. In 
Sec.~\ref{sec_unidirectional}, we examine separately the effects of periodic 
kicking of three different parameters in the Hamiltonian. We show that a 
rich variety of quasi-energy dispersions can be obtained depending on the 
kicking parameters. The effects of these dispersions on the wave packet 
dynamics will be shown in some particularly interesting cases. In
Sec.~\ref{sec_dl}, we present a general way of understanding the phenomenon
of dynamical localization. We end in Sec.~\ref{sec_concl} with a summary of 
our main results, possible experimental realizations of periodic kicking of 
the different parameters, and some directions for future work.

\begin{figure}[htb]
\includegraphics[width=8cm]{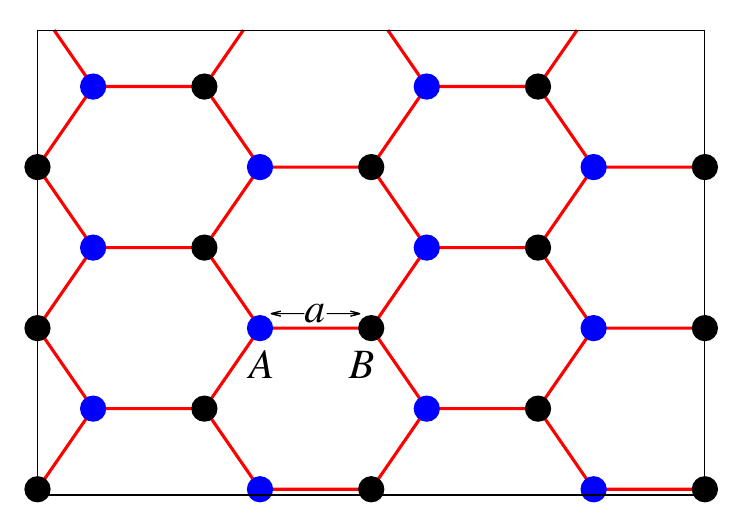}
\caption{The graphene lattice has a two-site unit cell; the sites are labeled 
$A$ and $B$ and are separated by a distance $a$.} \label{Fig:lat} \end{figure}

\section{Hamiltonian of graphene}
\label{sec_graphene}

We begin with a brief description of the band structure of graphene.
The Hamiltonian is given by a tight-binding model with nearest-neighbor 
hopping on a hexagonal lattice~\ct{neto09}
\be H ~=~ -~\ga ~\sum_{{\vec i}, {\vec j}} ~\sum_{s = \ua,\da}~
(c^\dg_{{\vec i},s} ~c_{{\vec j},s} ~+~ H. c.), \label{ham1} \ee
where the sum over ${\vec i}, ~{\vec j}$ goes over the nearest neighbors
(see Fig.~\ref{Fig:lat}),
the hopping amplitude $\ga \simeq 2.8 ~eV$, the nearest neighbor spacing
is $a \simeq 0.14 ~nm$, and $s$ denotes the spin component in, say, the 
$z$ direction. (We will henceforth set $\hbar =1$). Each unit cell of the 
hexagonal lattice consists of two sites; we denote the two sites, belonging to 
sublattices $A$ and $B$, by $a_\vn$ and $b_\vn$ respectively. We introduce the 
Pauli matrices $\vec \si$ with $\si^z = \pm 1$ denoting sites on the $A$ and 
$B$ sublattices respectively. The midpoint of a unit cell labeled as $\vn$ is 
located at $\vn ~=~ (3a/2) ~(n_1, (2n_2 + n_1)/\sqrt{3})$, where $n_1, ~n_2$ 
take integer values. The spanning vectors of the lattice are ${\vec M_1} = 
(3a/2) (1, 1/\sqrt{3})$ and ${\vec M_2} = (3a/2) (1, - 1/\sqrt{3})$. The 
reciprocal lattice vectors can be chosen to be ${\vec G_1} = (2\pi/3a)
(1/2, \sqrt{3}/2)$ and ${\vec G_2} = (2\pi/3a) (1/2, -\sqrt{3}/2)$.

In momentum space, the Hamiltonian in Eq.~\eqref{ham1} takes the form
\bea H_\vk &=& - \ga \left( \begin{array}{cc}
0 & 1 + e^{i \vk \cdot {\vec M_1}} + e^{i \vk \cdot {\vec M_2}} \\
1 + e^{-i \vk \cdot {\vec M_1}} + e^{-i \vk \cdot {\vec M_2}} & 0 \end{array}
\right) \non \\
&=& - \ga ~(g_\vk ~\si^x ~-~ h_\vk ~\si^y), \label{ham2} \eea
where 
\bea g_\vk &=& 1 + \cos (\vk \cdot {\vec M_1}) + \cos (\vk \cdot {\vec M_2}),
\non \\
h_\vk &=& \sin (\vk \cdot {\vec M_1}) + \sin (\vk \cdot {\vec M_2}). \eea
As is well-known, the energy dispersion is given by $\pm E_\vk$ where
\bea E_\vk &=& \ga \sqrt{g_\vk^2 ~+~ h_\vk^2} \label{ek} \\
&=& \ga [ 3 + 2 \cos (\sqrt{3} k_y a) + 4 \cos (\frac{\sqrt{3}
k_y a}{2}) \cos (\frac{3 k_x a}{2}) ]^{1/2}. \non \eea
The two bands touch each other at two inequivalent points; these are located 
at the wave vectors $\pm (0, 4\pi/(3\sqrt{3} a))$ and are called $\vec K$ and 
$\vec K'$. Around these points, the effective low-energy continuum theory of 
graphene electrons takes the form of a two-dimensional Dirac Hamiltonian with 
\be H_D ~=~ \sum_\vk \psi_\vk^\dg [ v_F ~( \si^x k_x ~+~ \tau^z \si^y k_y)]
\psi_\vk, \label{dirham1} \ee
where $v_F = 3\ga a/2$ is the Fermi velocity, $\tau^z = \pm 1$ at ${\vec K} ~
({\vec K'})$ respectively (these are called valleys), and $\psi_\vk 
\equiv \psi_\vk^{\si \tau s}$ denote eight-component electron
annihilation operators with the components corresponding to
sublattice ($\si$), valley ($\tau$), and spin ($s$) degrees of
freedom. Equation~\eqref{dirham1} is the Dirac Hamiltonian and the
dispersion is given by $\pm E_\vk = \pm v_F | \vk|$, with a
four-fold degeneracy due to the valley and spin degrees of freedom.
We will ignore the spin degree of freedom in this paper.

If we add a staggered potential of the form $f \si^z$, so that sites on the 
$A$ and $B$ sublattices have on-site energies $f$ and $-f$ respectively, the 
Dirac Hamiltonian gets a mass $f$ leading to the dispersion $\pm E_\vk = \pm 
\sqrt{ v_F^2 \vk^2 + f^2}$; this has a gap of $2|f|$ at $\vk = 0$.

In the next section we study what happens when various parameters
in the Hamiltonian are given $\de-$function kicks with a time period $T$.

\section{Floquet Hamiltonian for periodic kicking}
\label{sec_floquet}

We now apply $\de$-function kicks to the system, described by the
following term in the Hamiltonian
\be H_{kick} ~=~ \sum_\vn ~(\al_x \si^x + \al_y \si^y + \al_z \si^z) ~
\sum_{m=-\infty}^\infty ~\de (t - m T). \label{kick1} \ee
We assume that the periodic kicks are the same for all unit cells
$\vn$; hence they add the term 
\be H_{\vk,kick} ~=~ (\al_x \si^x + \al_y \si^y + \al_z \si^z) ~
\sum_{m=-\infty}^\infty ~\de (t - m T) \label{kick2} \ee
to the momentum space Hamiltonian in Eq.~\eqref{ham2}.

The Floquet (stroboscopic) operator $U_{XYZ}$ which evolves the system
through one time period $T$ is given by 
\beq U_{XYZ} ~=~ U_{\vk,kick} U_\vk ~=~ e^{-i\vec{\al}\cdot\vec{\si}} ~
e^{-i H_\vk T} \equiv e^{-i H_{XYZ} T} \label{uxyz} \eeq
in momentum space. This also defines for us the {\it effective} stroboscopic 
Hamiltonian $H_{XYZ}$. The eigenvalues of $H_{XYZ}$ can be found by using 
the following identity for Pauli spin matrices, 
\beq e^{ia(\hat{n}.\vec{\si})}e^{ib(\hat{m}.\vec{\si})}=e^{ic(\hat{k}.
\vec{\si})}, \eeq
where $\{a,b,c\}$ are scalars and $\{\hat{n}$, $\hat{m}$, $\hat{k}\}$ are 
unit vectors. Here $c$ and $\hat{k}$ can be found in terms of $a,b, \hat{n}$ 
and $\hat{m}$ using the following relations:
\bea \cos c &=& \cos a \cos b - \hat{n}\cdot\hat{m}\sin a\sin b, \non \\
\hat{k} &=& \frac{1}{\sin c}(\hat{n} \sin a \cos b + \hat{m} \sin b\cos a 
\non \\
&& ~~~~~~~- \hat{n}\times \hat{m} \sin a \sin b). \eea

Hence the eigenvalues of $H_{XYZ}$ are equal to $\pm \varepsilon$, where
the quasi-energy $\varepsilon$ is given by
\bea \varepsilon &=& \frac{1}{T}\arccos ~[ \cos \al \cos (E_\vk T) \non \\
&& ~~~~~~~~~~~~+ \frac{ \ga (\al_x g_\vk - \al_y h_\vk) }{\al E_\vk} \sin \al 
\sin (E_\vk T) ], \label{effedis} \eea
with $\al = \sqrt{\al^2_x+\al^2_y+\al^2_z}$. In the special case of only 
$\al_z \neq 0$, i.e., the kicking is applied to only the on-site energies of 
the graphene sublattice, the quasi-energy takes the particularly simple form 
$\varepsilon = \frac{1}{T}\arccos [ \cos \al_z \cos (E_\vk T)]$. 
We will discuss this in more detail below.

\section{Wave Packet Dynamics}
\label{sec_dynamics}

To corroborate our results from the analytical results presented in the later 
sections, we numerically study the time evolution of a wave packet on the
graphene lattice. We visualize the packet only at times which are integer
multiples of $T$, consistent with our understanding of the effective 
dispersion given by $\varepsilon$.
 
Consider a wave packet $\Psi$ in two spatial dimensions, with an initial 
momentum $\vk_o=(k_{ox},k_{oy})$ and a width $\si$. Namely,
\beq \Psi(\vec{r},t=0)=\frac{1}{\sqrt{2\pi\si^2}} ~\exp(-\frac{r^2}{4 \si^2}) ~
\exp(i \vk_o \cdot \vec{r}), \eeq
which is normalized such that $\int d\vec{r} |\Psi|^2=1$. The Fourier 
transform of $\Psi$ is given by,
\beq \Psi(\vk,t=0)= \sqrt{8 \pi \si^2} \exp[- \si^2 \{ (k_x-k_{ox})^2 + 
(k_y-k_{oy} )^2 \}]. \eeq 
such that $\frac{1}{(2\pi)^2}\int dk_x dk_y |\Psi(\vec{k})|^2 = 1$.

Since the system has translational symmetry even in the presence of kicking, 
we can study the time evolution of each $k$ mode separately. Thus 
a wave packet centered at $\vk_o$ can be evolved via $U_{XYZ}$ and then
Fourier transformed to the real space lattice. At each $\vk$ mode we have
a two-component vector associated with the occupancy of the graphene 
bands. Note that in the absence of any kicking, the upper and lower bands 
(positive and negative energies) have opposite velocities. Hence the wave 
packet movement will depend on the band which is initially occupied. In 
this work we always consider wave packets built out of the lower band
(negative energy). 

It may be useful to say a few words about our numerical procedure. We impose 
periodic boundary conditions in both real and momentum space. In momentum 
space, the Brillouin zone is a rhombus whose corners are at $(\pi /a) (-1/3, 
-1/\sqrt{3})$, $(\pi /a) (-1, 1/\sqrt{3})$, $(\pi /a) (1/3, 1/\sqrt{3})$, and 
$(\pi /a) (1, -1/\sqrt{3})$. If the number of unit cells is $N^2$ (the number 
of sites is $2 N^2$), the real space lies in a rhombus whose center lies at 
$(0,0)$ and corners are at $(3aN/2) (-1/2,-\sqrt{3}/2)$, $(3aN/2) (1/2,
-1/(2\sqrt{3}))$, $(3aN/2) (-1/2,1/(2\sqrt{3}))$, and $(3aN/2) (1/2,
\sqrt{3}/2)$. In Figs.~\ref{Fig:wavdispline} and \ref{Fig:DL}, we have taken 
$a =1$ and $N=40$. 

\section{Dispersion for unidirectional kicking}
\label{sec_unidirectional}

We now look at the effects of kicking in each of the three directions
in the subsections below. We will show that kicks along $\si^x$, $\si^y$ 
and $\si^z$ can have quite different effects.

\subsection{X-kicking:~ $\al_x \neq 0$, $\al_y=\al_z=0$}

We first consider the case when the kicking is applied only in the $\si^x$ 
direction. In this case, we find that the quasi-energy spectrum given in
Eq.~\eqref{effedis} is gapless (i.e., $\varepsilon = 0$), when $h_\vk =0$, 
$E_\vk = \ga |g_\vk|$, and $E_\vk T = \al = |\al_x|$. This leads to the 
equations
\bea 2 \cos(\frac{3k_x a}{2})\cos(\frac{\sqrt{3}k_ya}{2}) &=& 
\frac{\al_x}{\ga T} ~-~ 1, \non \\
2\sin(\frac{3k_x a}{2})\cos(\frac{\sqrt{3}k_ya}{2}) &=& 0. \eea
This gives the following gapless points
\bea \cos(\frac{\sqrt{3}k^g_ya}{2}) &=& \frac{1}{2} ~(\frac{\al_x}{\ga T} ~-~ 
1), \non \\
\sin(\frac{3k^g_xa}{2}) &=& 0. \eea
Clearly this can only be satisfied if $ -\ga T \leq \al_x \leq 3 \ga T$. We 
can also see that only $k^g_y$ can be modulated using $\al_x$. The low-energy 
dispersion about these gapless points can be found by expanding 
$\varepsilon$ as follows,
\bea \cos (\varepsilon T) &=& \cos \al_x \cos (E_\vk T) \non \\
&& + ~(1- \frac{h_\vk^2}{2 g_\vk^2}) ~\sin \al_x \sin (E_\vk T), \eea
which implies that
\beq \varepsilon^2 = \frac{1}{T^2} \left[ (\al_x-E_\vk T)^2 + 
\frac{\ga^2 T^2 h_\vk^2}{\al_x^2} \sin^2 \al_x \right]. \eeq
The effective velocities about the gapless points are given by
\bea v_x &=& \frac{3a}{2T} \sqrt{\frac{ (\al_x- \ga T)^2 \sin^2 \al_x}{
\al_x^2}}, \non \\
v_y &=& \frac{3a}{2T} \sqrt{\frac{3 \ga^2 T^2 + 2 \ga T \al_x -\al_x^2}{3}}. 
\eea

\begin{figure}[htb]
\centering
\includegraphics[width=8cm]{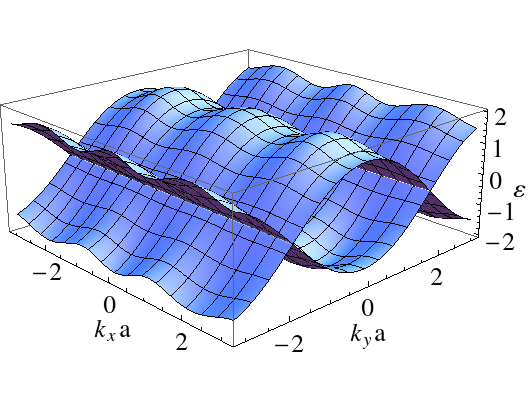}
\caption{Quasi-energy dispersion $\varepsilon$ (in units of $1/T$)
when $\al_x = \ga T = 1$ and $\al_y=\al_z=0$. 
There is a dispersionless line along the $k_x$ direction when $k_y a= \pm \pi /
\sqrt{3}$. A wave packet localized on this gapless line moves only in the $y$ 
direction (see Fig.~\ref{Fig:wavdispline}).} \label{Fig:displine} \end{figure}

\begin{figure*}[htb]
\centering
\includegraphics[width=12cm]{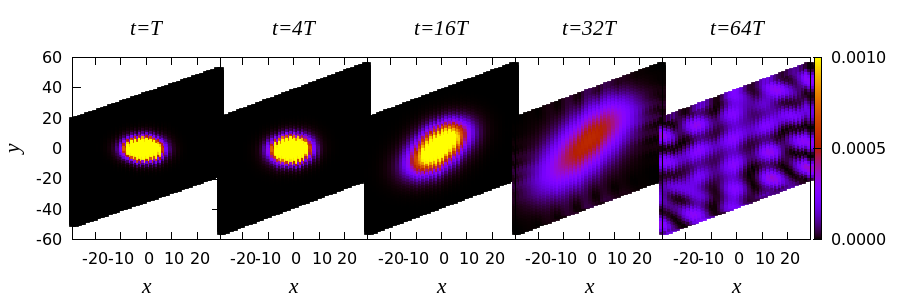}
\includegraphics[width=12cm]{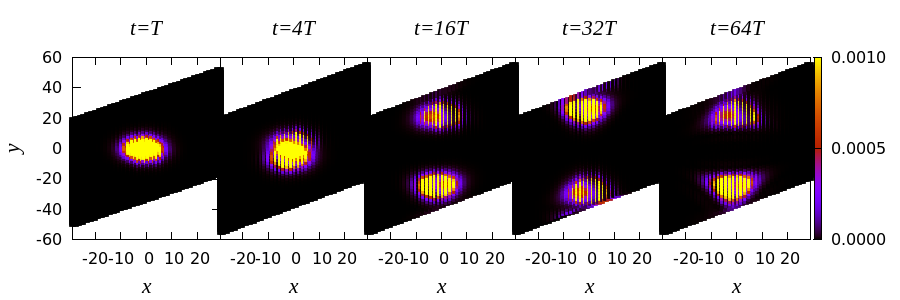}
\caption{Time evolution of a wave packet initially centered at $\vec{r}=(0,0)$,
at $t=T, ~4T, ~16T, ~32T, ~64T$ for no kicking (upper panel) and 
for $\al_x= \ga T = 1$ and $\al_y = \al_z = 0$
(lower panel), with $k_{ox}a=1$, $k_{oy}a=\frac{\pi}{\sqrt{3}}$ and $\si= 
\frac{10}{2\sqrt{2}}a$. The upper panel shows that in the absence of kicking 
the wave packet spreads out in both $x$ and $y$ directions. In the lower panel
we see that the wave packet only moves in the $y$ direction. This occurs 
because the wave packet is located at the dispersionless line in the $k_x$ 
direction (see Fig.~\ref{Fig:displine}).} \label{Fig:wavdispline} \end{figure*}

Interestingly, at the special value of $\al_x= \ga T$, we obtain a dispersion 
which is gapless along $k_x$ and disperses only along $k_y$ at $k_y=k^g_y$. 
This dispersionless line is shown in Fig.~\ref{Fig:displine}, where we have
chosen $\ga T = 1$. Therefore a wave packet which is centered at $k_{oy}=k^g_y$
will move only in the $y$ direction in real space in the presence of such a 
kicking. This is demonstrated in Fig.~\ref{Fig:wavdispline}. 

Another interesting case occurs as $\al_x$ is increased further. At $\al_x=3
\ga T$, we find that the two Dirac points merge at $(k_x^g , k_y^g) = (0,0)$ 
and we get a semi-Dirac dispersion there with $v_x= (a/T) \sin (3 \ga T)$ and 
$v_y=0$. The dispersion is linear in $k_x$ and quadratic in $k_y$, given by 
$\varepsilon=(3k^2_y a^2 \ga) /4$ for $k_x = 0$.

We would like to mention here that the merging of Dirac 
points in graphene resulting in a topological merging transition from a 
gapless to a gapped system has been studied extensively in recent years 
\ct{delplace11,kim12,koghee12,delplace13}. Further, the same phenomenon has 
been studied inside the gapless phase of the Kitaev model on the hexagonal 
lattice in Ref.~\onlinecite{bhattacharya08}.

\begin{figure}[htb]
\centering
\includegraphics[width=8cm]{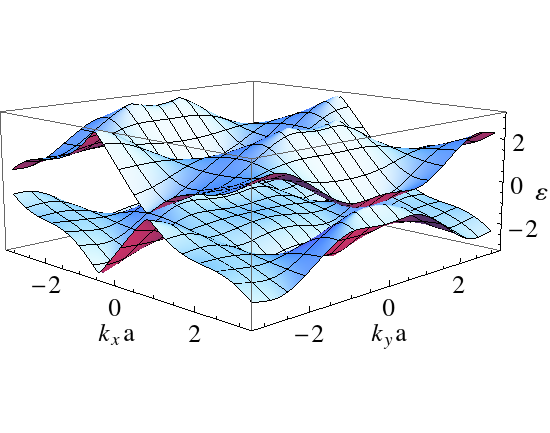}
\caption{Quasi-energy dispersion $\varepsilon$ (in units of $1/T$) when $\al_y=
\sqrt{3} \ga T = \sqrt{3}$ and $\al_x=\al_z=0$. While the dispersion is linear
in the $k_x$ direction, it is quadratic in the $k_y$ direction near the gapless
points (see text). This realizes a semi-Dirac dispersion.} \label{Fig:SD} 
\end{figure}

\subsection{Y-kicking:~ $\al_y \neq 0$, $\al_x=\al_z=0$}

Next we consider the case when the kicking is applied only in the $\si^y$ 
direction, i.e., $\al_y \neq 0$ and $\al_x=\al_z=0$. Substituting the above 
values in Eq.~\eqref{effedis}, the gapless points are determined by 
$g_\vk =0$, $E_\vk = \ga |h_\vk|$, and $E_\vk T= |\al_y|$. We then get
\bea 2 \cos(\frac{3k_x a}{2})\cos(\frac{\sqrt{3}k_ya}{2}) &=& -1, \non \\
2\sin(\frac{3k_x a}{2})\cos(\frac{\sqrt{3}k_ya}{2}) &=& -\frac{\al_y}{\ga T}. 
\eea
These can be solved to obtain gapless points at $(k^g_x, k^g_y)$, where
\bea \cos(\frac{\sqrt{3}k^g_ya}{2}) &=& - ~\frac{1}{2}\sqrt{1+ 
\frac{\al_y^2}{\ga^2 T^2}}, \non \\
\sin(\frac{3k^g_xa}{2}) &=& \frac{\al_y}{\sqrt{\ga^2 T^2+ \al^2_y}}. \eea
We see that a gapless point exists only if $|\al_y| \leq \sqrt{3} \ga T$. 
In contrast to the earlier case with only $\al_x \ne 0$, both $k^g_x$ and 
$k^g_y$ can be modulated here using $\al_y$. 

Next we obtain the effective dispersion about these gapless points. Similar to 
the previous section, the low-energy expansion leads to
\bea \cos(\varepsilon T) &=& \cos \al_y \cos (E_\vk T) \non \\
&& + ~(1- \frac{g_\vk^2}{2h_\vk^2}) ~\sin \al_y \sin (E_\vk T). \eea
Since the arguments of the both cosine terms are small, we expand both of them 
to obtain
\beq \varepsilon^2 = \frac{1}{T^2} \left[(\al_y-E_\vk T)^2 + 
\frac{\ga^2 T^2 g_\vk^2}{\al_y^2} \sin^2 \al_y \right]. \eeq
We see that the gapless points are Dirac-like and have anisotropic velocities 
given by
\beq v_{x}= \frac{3a}{2T} \sqrt{\ga^2 T^2 + \sin^2 \al_y}, \eeq
\beq v_y = \frac{a}{T} \sqrt{ \frac{3 \left(3\ga^2 T^2-\al_y^2\right) 
\left( \ga^2 T^2 \sin^2 \al_y + \al_y^4 \right)}{4 \al_y^2 \left(
\al_y^2+ \ga^2T^2 \right)}}. \eeq
Interestingly, at $|\al_y| =\sqrt{3} \ga T$, where the velocity $v_y$ vanishes,
we find a semi-Dirac dispersion which is linear in $k_x$ but quadratic in the 
$k_y$ direction; the low-energy dispersion in the $k_y$ direction is given by
\beq \varepsilon(k^g_x, k_y) = \frac{k_y^2 a^2}{T} ~\sqrt{\frac{3}{64} \left[ 
9 \ga^2 T^2 + \sin ^2 (\sqrt{3} \ga T) \right]}. \eeq
This is shown in Fig.~\ref{Fig:SD}, where we have chosen $\ga T = 1$.

\subsection{Z-kicking:~ $\al_z \neq 0$, $\al_x=\al_y=0$}

\begin{figure*}[htb]
\centering
\includegraphics[width=12cm]{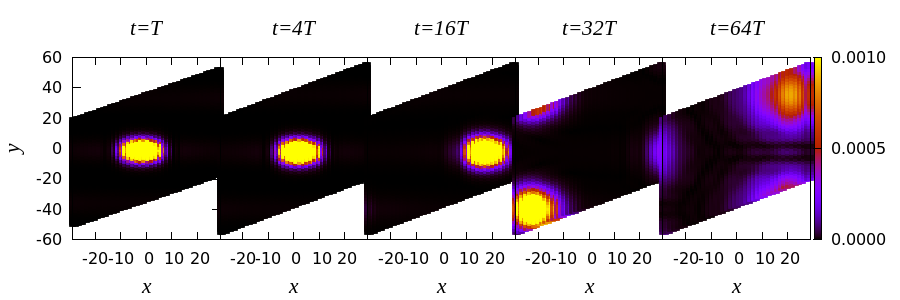}
\includegraphics[width=12cm]{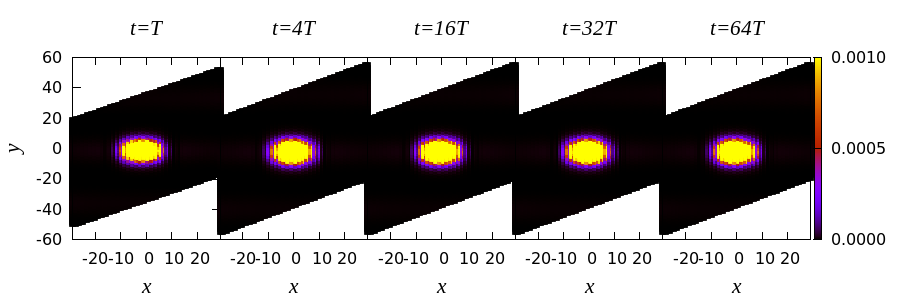}
\caption{Time evolution of a wave packet initially centered at $\vec{r}=(0,0)$,
at $t=T, ~4T,~16T, ~32T, ~64T$ for no kicking (upper panel) and for $\al_z=
\pi/2$ (lower panel), with $k_{ox}a=1$ and $k_{oy}a=0$ and $\si= \frac{10}{2
\sqrt{2}}a$. The upper panel shows that the wave packet evolves with a net 
velocity in the $x$ direction. The lower panel clearly demonstrates that the 
wave packet is localized. Notice that this happens in the absence of any 
disorder, so that translational symmetry is preserved. This is called 
dynamical localization.} \label{Fig:DL} \end{figure*}

Finally we consider a kicking which is applied in only the $\si^z$ direction;
this corresponds to applying a staggered potential on the $A$ and $B$ 
sublattices. The dispersion in this case is given by
\beq \varepsilon = \frac{1}{T}\arccos [ \cos \al_z \cos (E_\vk T)]. \eeq 
At the Dirac points, $E_\vk = 0$, we see that $\varepsilon = \al^z/T$; hence 
this kicking opens up a gap in the dispersion. Interestingly, at $\al_z = \pi /
2$, we find that $\varepsilon = \pi/ (2T)$, independent of the value of $\vk$.
Thus the dispersion becomes absolutely flat and therefore leads to dynamical 
localization. This is clearly shown in Fig.~\ref{Fig:DL} where a wave packet 
gets localized in real space. Note that this is happening even though the 
system has no disorder and has translational symmetry.

At all other values of $\al_z \ne n\pi$ (where $n$ is an integer), the 
original Dirac points at ${\vec K}$ and ${\vec K'}$ have a gap proportional
to $\al_z /T$ and a low-energy dispersion which is quadratic in $k$. The 
effective dispersion is given by 
\beq \varepsilon ~=~\frac{\al_z}{T} ~+~\frac{1}{2} ~\cot \al_z ~E^2_\vk ~T. 
\eeq

\section{Dynamical localization}
\label{sec_dl}

In this section we will present a general understanding of dynamical 
localization due to periodic kicking. Suppose that there is a 
time-independent Hamiltonian $H$ whose 
eigenstates and eigenvalues come in pairs, namely, $\psi_j$ and $\psi'_j$ 
with energies $E_j$ and $-E_j$. Let us assume that there is a unitary operator
$P$ which produces this transformation, namely, $P H P^{-1} = - H$, 
$P \psi_j = \psi'_j$ and $P \psi'_j = \psi_j$; hence $P^2 = I$. Then we can
show as follows that a periodic kick with $P$ will produce dynamical 
localization after two time periods. We begin with an eigenstate $\psi_j$ and
evolve it with the Hamiltonian $H$ for a time $T$; hence it picks up a 
phase $e^{-iE_jT}$. Then we kick it with $P$ which converts it to the
state $\psi'_j$. Upon evolving this with $H$ for a time $T$, it picks up the 
phase $e^{iE_jT}$; the two phases cancel each other exactly. Then another 
kick with $P$ brings it back to the state $\psi_j$. Thus any eigenstate 
$\psi_j$ will remain unchanged after a time $2T$. Since any state can be 
written as a superposition of eigenstates of $H$, we see that any state will 
remain the same after a time $2T$; this implies dynamical localization.

For any bipartite lattice with hopping only between sites belonging to 
different sublattices (graphene is a special case of this), we can see that an
operator $P$ which changes the phase of an eigenstate on only one sublattice 
by $-1$ produces another eigenstate with the opposite energy. A kick with a 
staggered potential of strength $\pi/2$ is precisely equivalent to such an 
operator $P$ (up to an overall phase of $i$). This explains why we observe 
dynamical localization when $\al_z = \pi/2$ and $\al_x = \al_y = 0$. In fact, 
dynamical localization will occur even if we consider a finite piece of 
graphene with an arbitrary size and shape. 

\section{Concluding remarks}
\label{sec_concl}

We have shown that applying $\de$-function kicks in different directions
in the sublattice space in graphene can lead to interesting physics 
including dispersionless lines in momentum space, semi-Dirac dispersion 
and even a completely flat dispersion. We have shown that these lead
to rich possibilities for the time evolution of wave packets, such
as motion along only one direction or a complete dynamical localization.
Given the widespread interest in graphene, the ability to tune its 
dispersion and obtain a range of dynamical behaviors should lead 
to applications in a variety of settings.

We note that a dynamical localization-to-delocalization transition has been 
observed in a quantum kicked rotor. Such a system is realized by placing 
cold atoms in a pulsed, far-detuned, standing wave, and the transition is
detected by measuring the number of zero velocity atoms under the influence 
of a quasi-periodic driving \ct{ringot00}.

We would like to mention possible experimental realizations of 
periodic driving of a graphene system.

\noi (i) If a metallic sheet carrying a uniform current in the $x$ direction
is placed parallel to the graphene (but displaced from it by some distance 
in the $z$ direction), this produces a constant magnetic field in the $y$ 
direction. The corresponding vector potential can be chosen to be in the $x$ 
direction with a magnitude which is linear in the $z$ coordinate. Hence it 
will be a constant vector in the graphene plane. This vector potential can be 
introduced as a Peierls phase in the nearest neighbor hoppings. If we now vary 
the current periodically in time, we will have a periodically varying hopping 
phase which cannot be removed by a global gauge transformation. This provides 
a possible experimental route to achieve the temporal driving in the $\si^x$ 
and $\si^y$ directions that we have studied in this paper.

\noi (ii) A kicking proportional to $\si^z$ can be experimentally set up as 
follows. h-BN (a form of boron nitride with a hexagonal lattice structure) and 
graphene have lattice spacings which are nearly equal; hence one of them can 
be placed on the other. The boron and nitrogen atoms exert different van der 
Waals forces on the two graphene carbon atoms in an unit cell, thus creating 
an effective sublattice potential \ct{jung15}. A periodic application of the 
pressure on these two layers (from the top and the bottom) should modulate 
the distance between the layers and thus lead to a periodic modulation of 
the sublattice potential. 

Finally, we point out some possible directions for future studies. One can 
study what kinds of edge states can be generated in graphene by periodic 
kicking of different kinds. In the absence of kicking, it is known that 
graphene has 
states on a zigzag edge but not on an armchair edge \ct{nakada96,kohmoto07}. 
It would be interesting to know if periodic driving can change this situation,
as is known to happen in the Kitaev model on the hexagonal lattice 
\ct{manisha14}. It would also be very interesting to analyze the effects of 
interactions in periodically kicked graphene. One of the central results of 
this paper is that at $\al_z = \pi/2$ the quasi-energy spectrum becomes 
completely dispersionless. Under these conditions any interaction energy scale
in the problem will be dominant due to quenching of the ``effective" kinetic 
energy. It can therefore be intriguing to understand the stroboscopic 
evolution of a many-particle state in such a system. The presence of 
a highly anisotropic Dirac dispersion and dispersionless lines in the 
spectrum may also produce exotic many-body phases in the presence of 
interactions. 

\section*{Acknowledgments}
We thank Arnab Das for interesting discussions. A.A. thanks CSIR, India for 
funding through a SRF fellowship. A.D. thanks DST, India for Project No. 
SB/S2/CMP-19/2013 and D.S. thanks DST, India for Project No. SR/S2/JCB-44/2010
for financial support.

\end{document}